%% file: ANW_disorder_paper_v6.tex
\documentclass[%
reprint,
superscriptaddress,
 amsmath,amssymb,
 aps,
]{revtex4-2}

\usepackage{graphicx}
\usepackage{dcolumn}
\usepackage{bm}
\usepackage{url}
\usepackage{verbatim}
\usepackage{float}
\usepackage{xcolor}
\usepackage{soul}
\usepackage[T1]{fontenc}
\usepackage{ragged2e}
\usepackage{textcomp}
\usepackage{lipsum}
\usepackage{xr}
\usepackage{ulem}
\newif\ifincludesm
\includesmtrue  

\definecolor{Magenta}{rgb}{0,0.85,0}

\definecolor{Blue}{rgb}{0,0,1}

\begin{document}

\preprint{APS/123-QED}



\title{Transport regimes of biphoton-state quantum walks in ordered and disordered arrays of nonlinear waveguides}


\author{Jefferson Delgado-Quesada}

\affiliation{Centro de Investigaci{\'o}n en Ciencia e Ingenier{\'i}a de Materiales, Universidad de Costa Rica, Costa Rica}

\affiliation{Escuela de F{\'i}sica, Universidad de Costa Rica, Costa Rica}

\author{Edgar A. Rojas-Gonz{\'a}lez}%

\email{edgar.rojasgonzalez@ucr.ac.cr}

\affiliation{Centro de Investigaci{\'o}n en Ciencia e Ingenier{\'i}a de Materiales, Universidad de Costa Rica, Costa Rica}

\affiliation{Escuela de F{\'i}sica, Universidad de Costa Rica, Costa Rica}

\date{\today}

\begin{abstract}
A quantum walk in an ordered medium exhibits ballistic propagation. A related process is the driven quantum walk, in which the number of walkers varies along the propagation. In this work, we show that a driven quantum walk of biphoton states in an array of nonlinear waveguides does not propagate ballistically, but instead presents two transport regimes: superballistic and superdiffusive. We also study the effects of off-diagonal and diagonal disorder. In addition to the observation of Anderson localization, both types of disorder lead to the disappearance of the superballistic regime.
\end{abstract}

\maketitle


\textit{Introduction}---Quantum walks (QWs) are the analogous to classical random walks when the effects of quantum mechanics are considered, such as superposition and interference \cite{Aharonov1993}. Their applications span from quantum computing \cite{Qiang2024} to biological systems \cite{Hoyer2010,Dubovitskii2025}. Integrated photonic platforms are ideal for studying quantum coherent transport phenomena and, in addition, are promising for the development of quantum technologies \cite{Wang2019,Pelucchi2021}. A key example is an array of waveguides, in which photons undergo a continuous time QW due to evanescent coupling between the waveguides \cite{Peruzzo2010}. The arrays of nonlinear waveguides (ANW) are of keen interest because, contrary to its linear counterpart, they are active elements able to increment the level of entanglement of an initial state \cite{Solntsev2014}. 
In a typical QW, the number of walkers is constant, whereas in a driven QW they can be created and annihilated along the propagation \cite{Hamilton2014}. 


For a given transport phenomenon that obeys a power-law scaling, one can define a root-mean-square displacement (standard deviation) $\sigma \propto z^\gamma$, with $z$ the evolution variable and $\gamma$ an exponent that characterizes the transport regime. That is, localized ($\gamma=0$), subdiffusive ($0<\gamma<1/2$), diffusive ($\gamma=1/2$), superdiffusive ($1/2<\gamma$<1), ballistic ($\gamma=1$) and superballistic ($\gamma>1$). The study of transport regimes of different phenomena is highly relevant. For example, superballistic growth has attracted significant attention \cite{Hufnagel2001,Zhang2012,Levi2012,Stuetzer2013,Nguyen2016,Gholami2017,Dong2024}, including the possibility of superballistic  diffusion of entanglement \cite{Fitzsimons2005}. Also, the absence of diffusion of a wave function in a disordered lattice is of particular interest, which is known as Anderson localization \cite{Anderson1958}. In a coupled system, there are two main kinds of disorder that can be considered: diagonal disorder \cite{Pendry1982a}---which is the one originally discussed in the seminal paper of Anderson \cite{Anderson1958}---and off-diagonal disorder \cite{Pendry1982}. In the context of ANWs, they are related to disorder in the effective propagation constants of the waveguides and the coupling constants, respectively.

The propagation is ballistic in an  ordered linear array \cite{Perets2008}, but the driven QW case is conceptually different \cite{Hamilton2014,Raymond2025}, and hence is expected to exhibit dissimilar behavior. The current literature about ANWs including disorder is scarce. As far as we know, there is one work related to Anderson localization \cite{Bai2016} and two associated with topological protection \cite{Doyle2022,Zecchetto2025}, and all of them deal exclusively with off-diagonal disorder. To the best of our knowledge, we present for the first time a study of the transport regimes of biphoton states in driven QWs, including the effects of both diagonal and off-diagonal disorder. Here, we address the case of walkers (photon pairs) generated via spontaneous parametric down-conversion (SPDC) \cite{Solntsev2014}.



\begin{figure}[b]
    \centering
    \includegraphics[width=0.75\columnwidth]{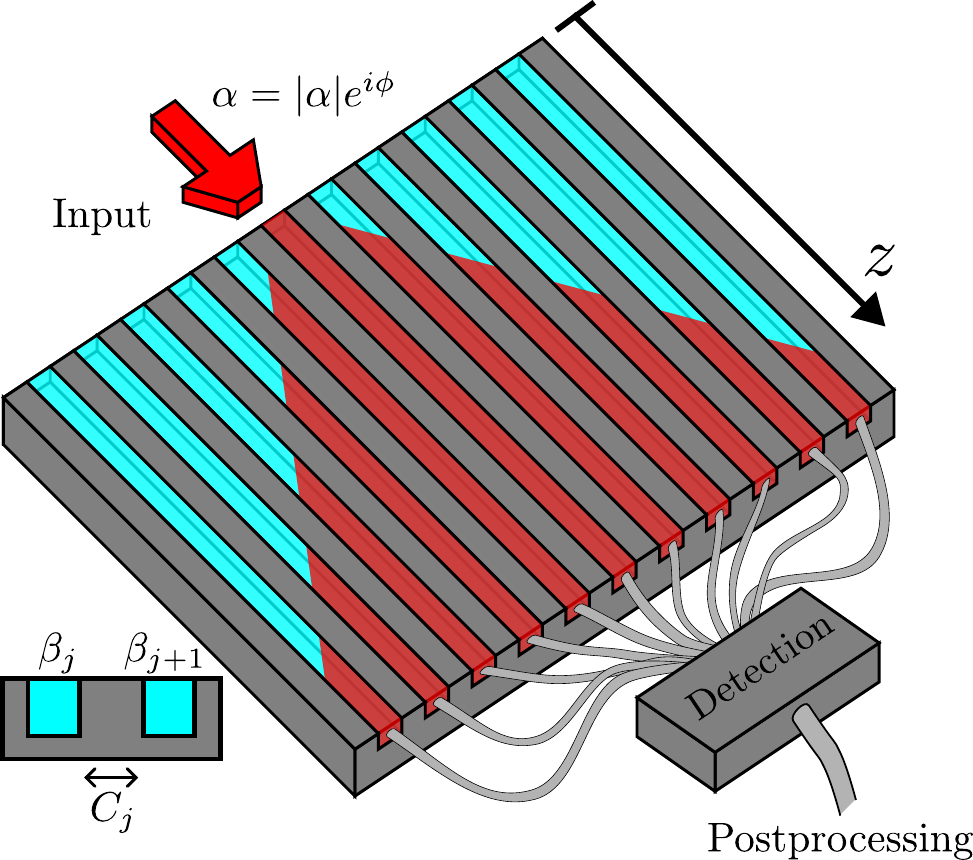}
    \caption{Schematic of a typical ANW with $N$ waveguides. Both linear coupling and nonlinear effects take place only within the ANW region, which begins at $z=0$. The effective propagation constant of waveguide $j$ is denoted by $\beta_j$ and $C_j$ corresponds to the linear coupling constant between waveguides $j$ and $j+1$. The output state, at the end of the ANW region, could be the input of a subsequent step.}
    \label{fig:ANW_figure}
\end{figure}

\textit{Characteristics of ANWs}---We hereby consider ANWs of $N$ waveguides with a quadratic nonlinearity determined by the second-order susceptibility $\chi^{(2)}$ (see Fig.~\ref{fig:ANW_figure}). Only the $p$-th waveguide is excited by a strong coherent pump field $\alpha=|\alpha|e^{i\phi}$, and the undepleted pump approximation is assumed to be valid. After injecting the array, a SPDC process takes place. Here, we circumscribe to the degenerate case, in which a pair of indistinguishable signal photons with frequency $\omega_\mathrm{s}=\omega_\mathrm{p}/2$ are generated from a pump photon with frequency $\omega_\mathrm{p}$. We focus on biphoton states, and thus assume that the injection intensity is low enough such that at most one nonlinear event takes place. As photons propagate in the ANW, evanescent coupling between waveguides also occurs. In this regard, we only consider nearest-neighbor coupling for the signal photons and negligible coupling for pump photons. That is, pump photons remain in their initial waveguide, which is a reasonable assumption since modal confinement increases with frequency \cite{Noda1981}.

The spatial evolution of photons in the ANW is governed by the momentum operator $\hat{M}$ \cite{Toren1994,Horoshko2022}, which takes the following form in the interaction picture \cite{Linares2008}

\begin{eqnarray} \label{Eq:1_M}
\hat{M} &=&\hbar \sum_{j=1}^{N-1} C_j \hat{A}_{j+1}\hat{A}_j^\dagger e^{i[\beta_{j+1}(\omega_\mathrm{s})-\beta_{j}(\omega_\mathrm{s})]z}\nonumber\\
& &+\hbar g \alpha (\hat{A}_p^\dagger)^2 e^{i[\tilde\beta_{p}-2\beta_{p}(\omega_\mathrm{s})]z}+\mathrm{h.c.},
\end{eqnarray}

with $\mathrm{h.c.}$ denoting the Hermitian conjugate, $z$ the direction of propagation along the ANW, $\beta_j(\omega_\mathrm{s})$ the effective propagation constant of the signal photons at the $j$-th waveguide, and here we consider  $\tilde\beta_{p}$ as a quantity that contains the effective propagation constant of the pump photons at the $p$-th waveguide and the required quasi-phase matching contribution (for example, due to periodic poling) to attain the condition $\tilde\beta_{p}-2\beta_p(\omega_\mathrm{s})=0$ within the ANW. In addition, $\hat{A}_j$ ($\hat{A}_j^\dagger$) corresponds to the monochromatic slowly-varying amplitude annihilation (creation) operator associated with the signal photons in the individual mode $j$ (the $j$-th waveguide), which satisfies the commutation relation $[\hat{A}_j(z,\omega),\hat{A}_{j'}^\dagger(z',\omega')]=\delta(z,z')\delta(\omega-\omega')\delta_{j,j'}$ \cite{BenAryeh1991}. The first term in Eq. \eqref{Eq:1_M} arises from the coupling between neighboring waveguides, quantified by $C_j$—the linear coupling constant between waveguides $j$ and $j+1$. The second term describes the SPDC process, whose strength is characterized by a nonlinear constant $g$, which is proportional to both $\chi^{(2)}$ and the overlap between the pump and signal fields.
In general, $g$ and $C_j$ depend on frequency, and both can be assumed to be real without affecting the underlying physics.

\textit{Methodology}---Our procedure is based on the analytic solution presented in Ref. \cite{DelgadoQuesada2025}. The generalization for nonidentical waveguides, necessary for diagonal disorder, can be found in Sec. \ref{sec:analytic_solution} in the Supplemental Material (SM). We define the normalized photon number as $n_k(z) = \langle \hat N_k\rangle / \sum_{j=1}^N\langle\hat N_j\rangle $, with $\hat N_k=\hat A_k^\dagger\hat A_k$ the number operator of the individual mode $k$. With this, we can calculate the standard deviation

\begin{equation}
    \sigma(z) = \sqrt{\sum_{k=1}^Nk^2n_k-\left(\sum_{k=1}^Nkn_k\right)^2} \label{Eq_sigma}.
\end{equation}

We also define

\begin{equation}
\gamma(z)=\frac{\mathrm{d}\log\sigma}{\mathrm{d}\log z}, \label{Eq_gamma}
\end{equation}
which corresponds to the exponent of the standard deviation when this is a power function---that is, $\sigma\propto z^\gamma$. If $\sigma$ is not a pure power function, this would manifest as a $z$ dependency in $\gamma$.



We hereby consider disorder as a random variation in a given parameter. In this case, diagonal (off-diagonal) disorder is associated with variations in $\beta_j$ ($C_j$). We model disorder by letting $\beta_j(\omega_\mathrm{s})$ and $C_j$ belong to the following uniform distributions


\begin{eqnarray}
    \beta_j(\omega_\mathrm{s}) &\in [\beta_0 - \kappa_\beta\Delta\beta,\beta_0 + \kappa_\beta\Delta\beta], \label{Eq_UD_betaj}\\
    C_j & \in [C_0 - \kappa_C\Delta C, C_0 + \kappa_C\Delta C], \label{UD_Cj}
\end{eqnarray}
where $\beta_0$ ($C_0$) and $\Delta\beta$ ($\Delta C$) represent, respectively, the average of the random variable $\beta_j(\omega_\mathrm{s})$ ($C_j$) and its maximum variation---which were chosen as $\Delta C=0.9 C_0$ and $\Delta \beta=3 C_0$. The parameters $\kappa_\beta$ and $\kappa_C$, which range from 0 to 1, quantify the diagonal and off-diagonal disorder strength, respectively. Note that the results are general for any choice of $C_0$ and $\beta_0$ because, if $\Delta \beta$ is expressed in terms of $C_0$, then $C_0$ and $z$ are present in the solution solely as the dimensionless propagation parameter $C_0z$. Also, $\beta_0$ only adds a global phase, which has no impact in the physics.
All the results presented in this work correspond to the average of 200 simulations, which were carried out in Python by numerical implementation of the analytical solutions. Details of the simulations as well as the scripts can be found in Ref. \cite{Repository_ANW_transport}.

\begin{figure}[t]
    \centering
    \includegraphics[width=\columnwidth]{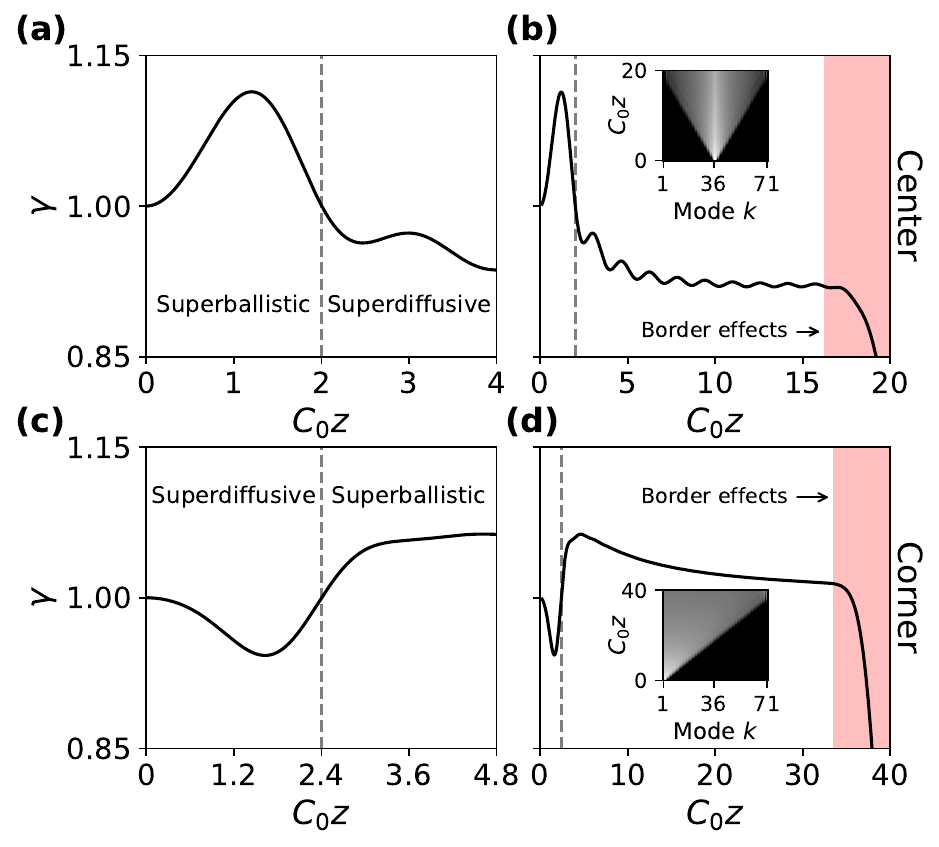}
    \caption{Parameter $\gamma$ as a function of the propagation distance $C_0z$ for a 71-waveguide array. The upper (lower) row corresponds to when the center (corner) waveguide is pumped. The left (right) column shows a small (large) propagation distance. For each plot, the vertical dashed gray line indicates the transition between a superballistic and a superdiffusive transport regime. The red-shaded regions in the right column correspond to where border effects are significant. The insets on the right column depict the normalized photon number along the propagation (the lighter the color, the larger the values).}
    \label{fig:ordered_propagation}
\end{figure}

\textit{Ordered array}---We first study a disorder-free array (with $N=71$) when only the center waveguide is pumped. In such case, the border effects become relevant for $C_0 z \gtrsim 16$ (see upper row in Fig.~\ref{fig:ordered_propagation}). These manifest as a decay in $\gamma$, which correlates with an increase of the normalized photon number at one of the corners ($n_1$ or $n_N$), as shown in Fig.~\ref{fig:size_effect_center} in SM. If one wants to study longer propagation distances, a larger array is required to avoid these effects. The variation of $\gamma$ with the propagation is depicted in
Figs.~\ref{fig:ordered_propagation}(a) and \ref{fig:ordered_propagation}(b). Although the propagation plot evokes a linear growth of the variance $\sigma$, see inset in Fig.~\ref{fig:ordered_propagation}(b), the transport is not purely ballistic because $\gamma$ differs from one for most of  the values of $C_0z$. In fact, there are two distinct regimes: superballistic propagation ($\gamma>1$) for $C_0z<2.005$ and superdiffusion ($1/2<\gamma<1$) for $C_0z>2.005$. In the former, $\gamma$ increases up to 1.114 at $C_0z=1.203$. At large propagation distances, see Fig.~\ref{fig:ordered_propagation}(b), $\gamma$ seems to reach a steady value. For instance, in an array with $N=1000$, we get $\gamma=0.936$ at $C_0z=200$ (Fig. \ref{fig:size_effect_center} in SM), which suggests that the propagation remains in a superdiffusive regime for large $z$. The $z$ dependence of $\gamma$ within the region without border effects does not change with the size of the array, see Fig. \ref{fig:size_effect_center} in SM. As a result, for the case of biphoton states, a disorder-free ANW will always exhibit a transition from superballistic to superdiffusive propagation at $C_0z=2.005$ as long as $N\gtrsim14$. In fact, the physical position $z$ where the transition occurs can be customized by tuning $C_0$. 

When a waveguide other than the center one is pumped, it is possible to observe the effects of both the first and second border encountered by the light (see Fig.~\ref{fig:border_effect} in SM). As mentioned before, this manifests as a significant decrease in $\gamma$, see Fig.~\ref{fig:size_effect_corner} in SM corresponding to an injection in the corner. Particularly, we also found relevant to analyze an injection in a corner of an ANW to study the transport process in a system that resembles a semi-infinite medium.
In this case, depicted in the lower row of Fig.~\ref{fig:ordered_propagation}, the propagation also resembles a linear growth of the variance, see inset in Fig.~\ref{fig:ordered_propagation}(d). Interestingly, $\gamma$ shows the opposite behavior compared to that observed when the center waveguide is pumped. That is, superdiffusion in the region $C_0z<2.404$ and superballistic propagation for larger distances. In the former, $\gamma$ reaches a minimum of $0.943$ at $C_0z=1.630$, while it presents a maximum of $1.063$ at $C_0z=4.619$ in the superballistic domain. In Fig.~\ref{fig:ordered_propagation}(d), the propagation remains in a superballistic regime for larger values of $C_0 z$. Nonetheless, it reaches $\gamma=1.001$ at $C_0z=400$ for an array of size $N=1000$ (see Fig. \ref{fig:size_effect_corner} in SM), suggesting that the propagation tends to a ballistic regime at large distances. 
We emphasize that the transition between transport regimes (at the $C_0 z$ values reported here) is a general result for biphoton states in  any ordered ANW injected in the center or in one corner provided that no border effects are relevant.

\begin{figure*}
    \centering
    \includegraphics[width=\linewidth]{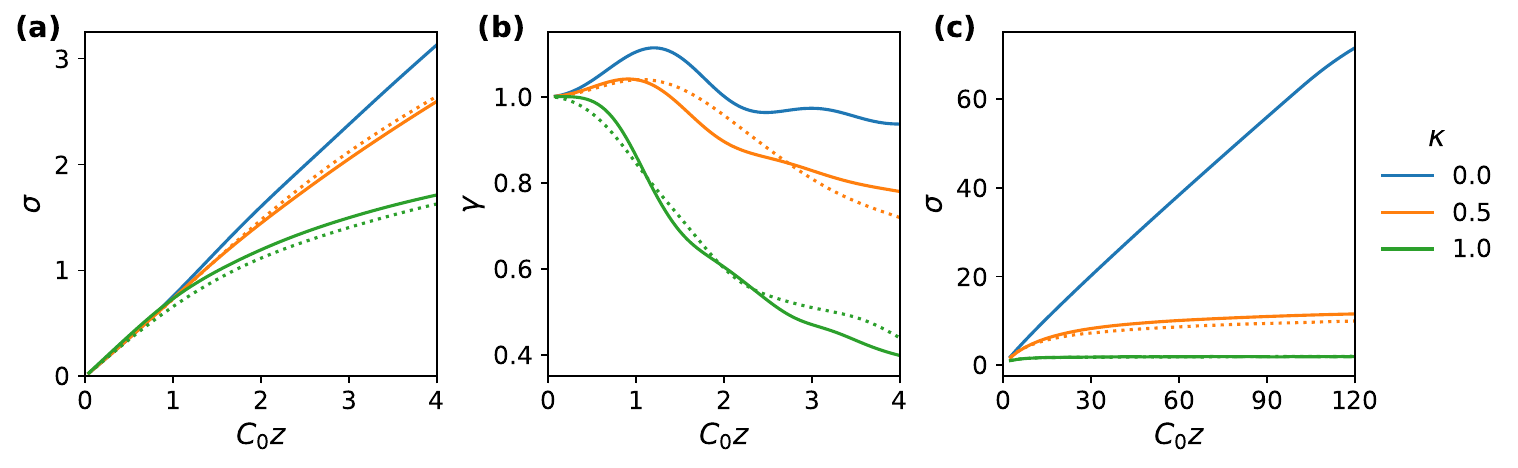}
    \caption{Standard deviation $\sigma$, (a), and $\gamma$, (b), at small propagation distances for an injection in the center of a 51-waveguide array when either off-diagonal (solid lines) or diagonal disorder (dashed lines) is introduced. Panel (c) depicts large propagation distances for a 401-waveguide ANW injected in the center. The color of each line represents the disorder strength $\kappa$ ($\kappa_C$ or $\kappa_\beta$). In (c), the case with $\kappa=0$ presents appreciable border effects for $C_0z \gtrsim 100$, see Fig. \ref{fig:large_prop}(b) in the End Matter.}
    \label{fig:center_pump_MT}
\end{figure*}

\textit{Disordered array}---In this work, we analyzed the cases of injections in the center or corner waveguides with diagonal and off-diagonal disorder. In general, when disorder is added to the system, $\sigma$ flattens and $\gamma$ decreases with propagation, as depicted in Fig. \ref{fig:small_prop} in the End Matter. The particular example of an injection in the center, for few levels of disorder, is shown in Fig. \ref{fig:center_pump_MT}.  These effects are more prominent with increasing disorder strength. Indeed, this behavior is a hallmark of Anderson localization, which can be clearly observed when a large propagation range is portrayed, see Fig. \ref{fig:center_pump_MT}(c) and Fig. \ref{fig:large_prop} in the End Matter.

\begin{figure}[h]
    \centering
    \includegraphics[width=\columnwidth]{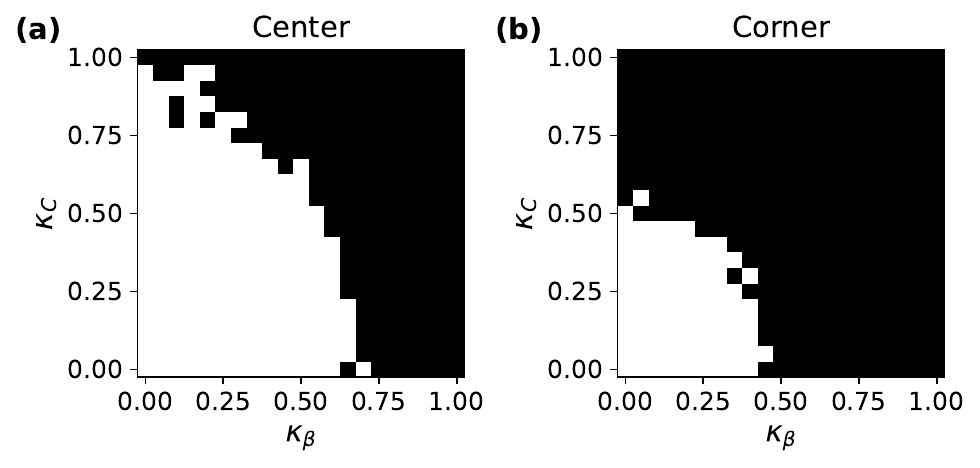}
    \caption{Color map showing the combinations of the disorder strengths, $\kappa_c$ and $\kappa_\beta$, for which the superballistic regime is present (white pixels) or absent (black pixels). The left (right) column corresponds to an injection in the center (corner) of a 51-waveguide array.}
    \label{fig:transition}
\end{figure}

When both kinds of disorder are present simultaneously, the superballistic regime disappears for combinations of $\kappa_C$ and $\kappa_\beta$ corresponding to high disorder strengths, as shown in Fig. \ref{fig:transition}. This point is illustrated in Figure \ref{fig:center_pump_MT} for the case of either only off-diagonal or diagonal disorder. Indeed, $\sigma$ flattens with $\kappa_C$ or $\kappa_\beta$, see Figs. \ref{fig:center_pump_MT}(a) and \ref{fig:center_pump_MT}(c), and the superballistic region (with $\gamma>1$) decreases with the disorder strength until it vanishes completely, see Fig. \ref{fig:center_pump_MT}(b). In fact, the same qualitative behavior is observed when either off-diagonal or diagonal disorder is present in an ANW injected either in the center or in the corner (see Fig. \ref{fig:small_prop} in the End Matter). We report a peculiar feature observed in an ANW injected in the center with the presence of diagonal disorder. That is, a small initial increase of $\sigma$ with the disorder strength $\kappa_\beta$, see Fig. \ref{fig:sigma_vs_kappa} in the End Matter.

As mentioned before, the results discussed so far do not depend on the specific values of $\beta_0$ and $C_0$. Typical experimental figures of $C_0$ for the relevant case of signal photons in the telecom about $1560\, \mathrm{nm}$ range in the order of $250-2500\, \mathrm{m}^{-1}$ \cite{Barral2021,Peruzzo2010,Raymond2025}. Also, $\beta_0=7.85 \times 10^6\, \mathrm{m}^{-1}$ assuming an effective refractive index of 1.95, which is reasonable because, for example, the bulk refractive index of lithium niobate
at $1560\, \mathrm{nm}$ is of about 2.21 (2.13) for the ordinary (extraordinary) axis \cite{Zelmon1997}. Indeed, $\beta_0 \sim 10^{7}\, \mathrm{m}^{-1}$ for usual effective refractive indexes of the order of few units. The chosen maximum variations $\Delta C=0.9 C_0$ and $\Delta \beta=3 C_0$ led to quantitatively similar effects in $\sigma$ (for example, in terms of the flattening of $\sigma$ with the propagation), see Fig. \ref{fig:center_pump_MT} and Figs. \ref{fig:small_prop} and \ref{fig:large_prop} in the End Matter. Despite $\Delta \beta>\Delta C$, compared to the mean values we have $\Delta C/C_0 = 0.9$ and $\Delta \beta/\beta_0 \sim 10^{-4}-10^{-3}$. Thus, in terms of percentages with respect to the corresponding mean values, for typical experimental figures the system is more sensitive to a random variation of the effective propagation constants (diagonal disorder) compared to that of the linear coupling constants (off-diagonal disorder).


\textit{Conclusion}---We characterized the transport regimes of biphoton states in ANWs. In an ordered homogeneous array, the process was not purely ballistic. Particularly, a transition was observed between the superballistic and superdiffusive regimes at low propagation distances. In fact, this transition occurs always at the same value of $C_0 z$ regardless of the characteristics of the array, provided that border effects can be neglected. At large propagation distances the system remains in a superdiffusive regime for an injection in the center, while it tends to a ballistic propagation for an injection in the corner waveguide. Indeed, the existence of the superballistic regime is a notable feature because it appears naturally even in an ordered homogeneous array (without the need for special coupling patterns in the lattice). With the introduction of disorder, Anderson localization effects were observed at large propagation distances. Moreover, the propagation region presenting superballistic transport shrinks (and even disappears) with the disorder strength. Lastly, for typical experimental figures the effect of disorder on the system is larger in the case of diagonal disorder. 
In future work, it would be valuable to explore the connection between the findings presented here and the evolution in the linear supermode basis of the array \cite{Kapon1984,Barral2020} and that of the correlation matrix \cite{Peruzzo2010}. The results obtained here are relevant for the engineering of integrated photonic platforms based on ANWs. Also, the study of disorder in ANWs sheds light on its effects and even its potential as a design parameter in integrated photonics applications. 
In addition, it is relevant for understanding the repercussions of random errors inherent in a manufacturing process.

\textit{Acknowledgments}---E.A.R.-G. and J.D.-Q. acknowledge support from the
Vice-rectory for Research at the University of Costa Rica (Project no. C6122).

\textit{Data availability}---The data that support the findings of this article are openly available \cite{Repository_ANW_transport}.

\bibliography{ANW_disorder_paper}

\newpage

\onecolumngrid

\section*{End matter}

\begin{figure*}[h]
    \centering
    \includegraphics[width=\textwidth]{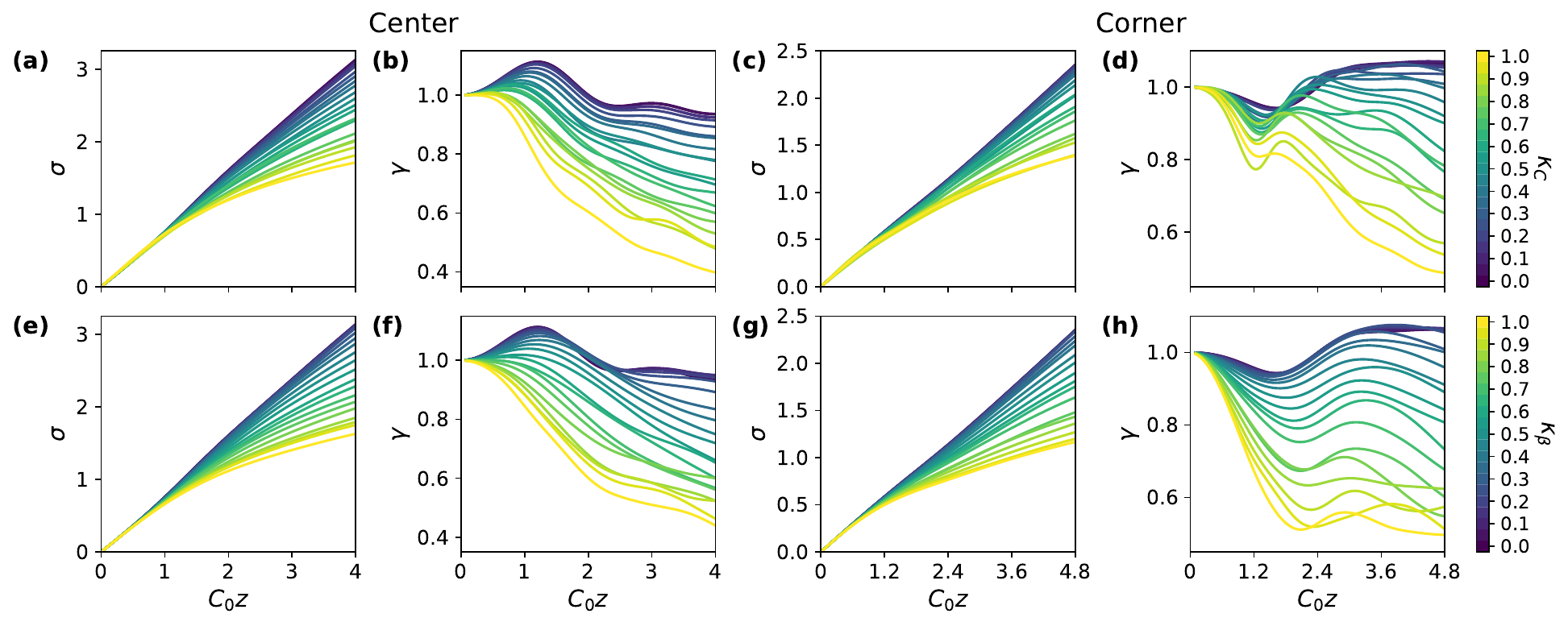}
    \caption{Standard deviation $\sigma$ and parameter $\gamma$ for small propagation distances when the center---(a), (b), (e), and (f)---or the corner waveguide---(c), (d), (g), and (h)---of a 51-waveguide array is pumped. The upper (lower) row corresponds to only off-diagonal (diagonal) disorder. The color of each line represents the strength of off-diagonal (diagonal) disorder, which is quantified by $\kappa_C$ ($\kappa_\beta$).}
    \label{fig:small_prop}
\end{figure*}

\textit{Appendix A: Effect of disorder for small propagation distances}---Figure \ref{fig:small_prop} depicts the standard deviation $\sigma$ and $\gamma$ as a function of the propagation distance for different diagonal and off-diagonal disorder strengths for an injection in the center and corner waveguide. At large, $\sigma$, and $\gamma$ show the tendency to flatten, and decrease with the disorder strength, respectively.

\begin{figure*}[h]
    \centering
    \includegraphics[width=\textwidth]{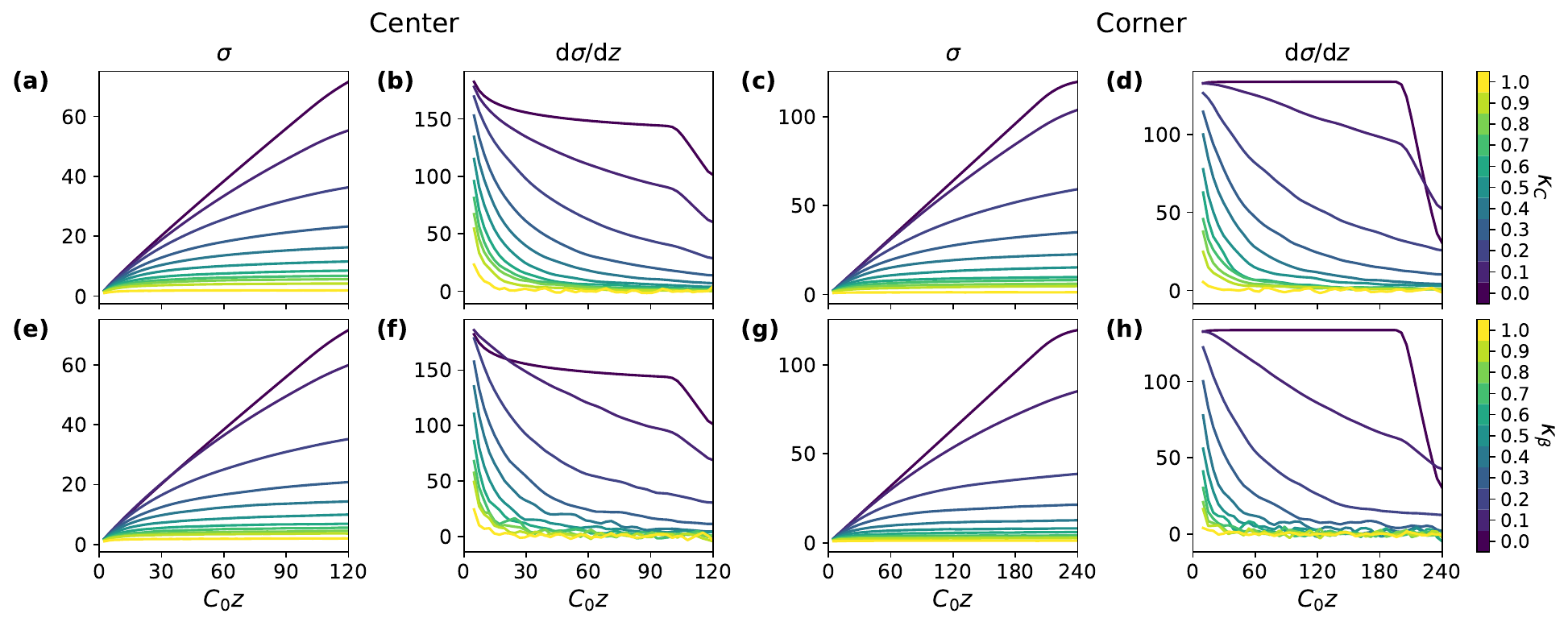}
    \caption{Standard deviation $\sigma$ (first and third columns) and its derivative $\mathrm{d}\sigma/\mathrm{d}z$ (second and fourth columns) as a function of propagation for different disorder strengths in an ANW of $N=401$ when the center---(a), (b), (e), and (f)---or the corner waveguide---(c), (d), (g), and (h)---is injected. The off-diagonal, and diagonal disorder cases are depicted in the first, and second row, respectively. The kinks observed in $\mathrm{d}\sigma/\mathrm{d}z$ for low disorder strengths are due to border effects.}
    \label{fig:large_prop}
\end{figure*}

\textit{Appendix B: Effect of disorder for large propagation distances}---Figure \ref{fig:large_prop} depicts the standard deviation $\sigma$ in the presence of diagonal and off-diagonal disorder for the cases of an injection in the center and corner waveguide. For the portrayed propagation distances, it is possible to observe clear signs of Anderson localization effects for the highest disorder strength values. That is, a complete flattening of $\sigma$, which is consistent with a derivative $\mathrm{d}\sigma/\mathrm{d}z$ that tends to zero, as shown in the second and fourth column of Fig. \ref{fig:large_prop}. Note that for some of the depicted disorder strength values a larger propagation range is required for $\sigma$ to completely flatten.

\begin{figure*}[h]
    \centering
    \includegraphics[width=0.7\textwidth]{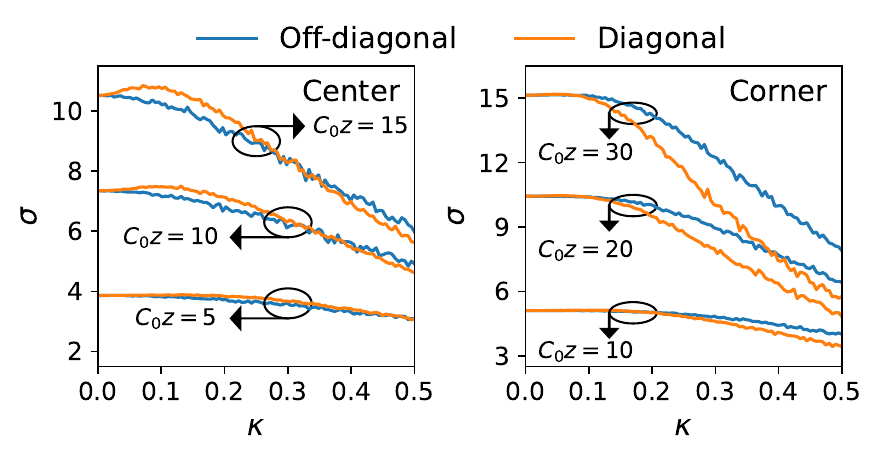}
    \caption{Standard deviation $\sigma$ as a function of disorder strength $\kappa$ ($k_C$ or $\kappa_\beta$). The left (right) column depicts $\sigma$ at the values $C_0z=\{5, 10, 10\}$ ($C_0z=\{10, 20, 30\}$) for an injection in the center (corner) of a 51-waveguide array. The blue, and orange lines represent, respectively, only off-diagonal, or diagonal disorder. For the case of an injection in the corner waveguide, the values of $C_0z$ were doubled since it takes a longer distance to reach similar values of $\sigma$ compared to an injection in the center waveguide.}
    \label{fig:sigma_vs_kappa}
\end{figure*}

\textit{Appendix C: Standard deviation as a function of disorder strength}---The dependence of $\sigma$ with the disorder strength at some fixed propagation distances is depicted in Fig. \ref{fig:sigma_vs_kappa}. Notably, a peculiar feature is observed in an ANW with diagonal disorder injected in the center---that is, a small initial increase in $\sigma$ with the disorder strength, see Fig. \ref{fig:sigma_vs_kappa}(a). Interestingly, this aspect disappears when only off-diagonal disorder is present. Also, for an injection in the corner, such an increase in $\sigma$ is not observed for either type of disorder, see Fig. \ref{fig:sigma_vs_kappa}(b). This feature seems to hold for any value of $C_0z$ provided that border effects are absent, becoming more pronounced for larger propagation distances.

\clearpage

\ifincludesm
\input{supplementary-materials_v2}  
\fi

\end{document}

%% file: supplementary-materials_v2.tex
\onecolumngrid
\setcounter{equation}{0}
\renewcommand{\theequation}{S\arabic{equation}}

\renewcommand{\appendixname}{}
\renewcommand{\thesubsection}{\Roman{subsection}}

\setcounter{figure}{0}
\renewcommand{\thefigure}{S\arabic{figure}}

\section*{Supplemental Material}

\subsection{Generalization of the analytic solution for an array with nonidentical waveguides}
\label{sec:analytic_solution}

For a nonlinear array, the momentum operator in the Schrödinger picture is given by

\begin{align}
    \hat M &= \hbar\sum_{j=1}^{N}\beta_j(\omega_\mathrm{s})\hat A_j^\dagger\hat A_j + \hbar\sum_{j=1}^{N}\beta_j(\omega_\mathrm{p})\hat A_{\mathrm{p}j}^\dagger\hat A_{\mathrm{p}j}+\left(\hbar \sum_{j=1}^{N-1} C_j \hat{A}_{j+1}^\dagger\hat{A}_j + \hbar g\sum_{j=1}^N \hat A_{\mathrm{p}j}(\hat{A}_j^\dagger)^2 +\mathrm{h.c.}\right) \label{Eq_SM_M}
\end{align}
where $\hat{A}_{\mathrm{p}j}$ ($\hat{A}_{\mathrm{p}j}^\dagger$) corresponds to the annihilation (creation) operator of a pump photon in the waveguide $j$. The remaining quantities are defined in the main text. The operator $\hat{M}$ in Eq. \eqref{Eq_SM_M} can be conveniently written as

\begin{equation}
    \hat{M} = \hat{M}_{\mathrm{P}}+\hat{M}_{\mathrm{L}}+\hat{M}_{\mathrm{NL}},
\end{equation}

with

\begin{align}
   \hat{M}_\mathrm{P}&=\hbar\sum_{j=1}^{N}\beta_j(\omega_\mathrm{p})\hat A_{\mathrm{p}j}^\dagger\hat A_{\mathrm{p}j},\\
   \hat{M}_\mathrm{L}&= \hbar\sum_{j=1}^{N}\beta_j(\omega_\mathrm{s})\hat A_j^\dagger\hat A_j+\left(\hbar \sum_{j=1}^{N-1} C_j \hat{A}_{j+1}^\dagger\hat{A}_j+\mathrm{h.c.} \right), \label{Eq_SM_ML}\\  \hat{M}_\mathrm{NL}&=\hbar g\sum_{j=1}^N \hat A_{\mathrm{p}j}(\hat{A}_j^\dagger)^2 +\mathrm{h.c.}. \label{Eq_SM_MNL}
\end{align}

Let us define the supermode basis $\{\hat{b}_j\}$ as the one that diagonalizes $\hat{M}_\mathrm{L}$ in Eq. \eqref{Eq_SM_ML} as follows

\begin{equation}
\hat{M}_\mathrm{L}= \hbar \sum^N_{j=1} \lambda_j \hat{b}^{\dagger}_j \hat{b}_j,
\end{equation}

with $\{\lambda_j\}$ the eigenvalues of the matrix

\begin{equation}
\label{Eq:5_L_form}
\Omega=\begin{pmatrix}
\beta_1(\omega_\mathrm{s})   & C_1 &         &        &         \\
C_1 &  \beta_2(\omega_\mathrm{s})  & C_2     &        &         \\
    & C_2 & \ddots  & \ddots &         \\ 
    &     & \ddots  & \ddots & C_{N-1} \\ 
    &     &         &  C_{N-1}   & \beta_N(\omega_\mathrm{s})   \\ 
\end{pmatrix}.
\end{equation}

Here, the eigenvalues are real because the matrix $\Omega$ is real and symmetric, and their labeling is defined such that $\lambda_1>\dots>\lambda_k>\dots>\lambda_N$.

The relation between the individual $\{\hat{A}_j\}$ and the supermode $\{\hat{b}_j\}$ bases is given by the transformation

\begin{equation}
b=S A,
\end{equation}

with $b=(\hat{b}_1,\dots,\hat{b}_N)^T$, $A=(\hat{A}_1,\dots,\hat{A}_N)^T$ and 

\begin{equation}
S = \begin{pmatrix}
\vec{v}_1^{\hspace{0.1cm}T}\\
\vdots\\
\vec{v}_N^{\hspace{0.1cm}T}
\end{pmatrix},
\end{equation}

with $\vec{v}_k$ the eigenvector of the matrix $\Omega$ associated with the eigenvalue $\lambda_k$.

It is convenient to switch to the interaction picture, in which the corresponding nonlinear term  of the momentum operator $\hat{M}^{(\mathrm{I})}_\mathrm{NL}$ is given by

\begin{equation}
 \hat{M}^{\mathrm{(I)}}_\mathrm{NL}(z)  = \hbar g\sum^N_{n,m=1} \sum^N_{j=1}  S_{nj} S_{mj} \hat{A}^{\mathrm{(I)}}_{\mathrm{p}j} \hat{b}^{\mathrm{(I)}\dagger}_n \hat{b}^{\mathrm{(I)}\dagger}_m + \mathrm{h.c.},
\end{equation}

with the subscript $(\mathrm{I})$ denoting that an operator is in the interaction picture. In addition, we have the following relations

\begin{align}
\hat{b}_n &= \hat{b}^{\mathrm{(I)}}_ne^{-i\lambda_n z}, \\
\hat{A}_{\mathrm{p}j} &= \hat{A}^{\mathrm{(I)}}_{\mathrm{p}j}e^{-i\beta_j(\omega_\mathrm{p})z}.
\end{align}

Thus, in terms of $\hat{b}_n$ and $\hat{A}_{\mathrm{p}j}$, $\hat{M}^{\mathrm{(I)}}_\mathrm{NL}(z)$ can be expressed as follows

\begin{align}
    \hat{M}^{\mathrm{(I)}}_\mathrm{NL}(z)  =& \hbar g\sum^N_{n,m,j=1} e^{i\beta_j(\omega_\mathrm{p})z} e^{-i(\lambda_n+\lambda_m)z}S_{nj} S_{mj} \hat{A}_{\mathrm{p}j} \hat{b}^{\dagger}_n \hat{b}^{\dagger}_m   + \mathrm{h.c.} \\
    =& \hbar g||\vec{\alpha}||\sum^N_{n,m,j=1} \eta_je^{i[\beta_j(\omega_\mathrm{p})-(\lambda_n+\lambda_m)]z}S_{nj} S_{mj} \hat{b}^{\dagger}_n \hat{b}^{\dagger}_m   + \mathrm{h.c.},
\end{align}


where a classical approximation for the pump was used in the last step. That is, $\hat A_{\mathrm{p}j}\rightarrow \alpha_j=||\vec\alpha||\eta_j$, with $\alpha_j$ a complex number corresponding to the pump field injected in waveguide $j$, $\vec{\alpha}=(\alpha_1,\dots, \alpha_N)^T$, and $\eta_j$ the $j$th component of the unit vector $\hat{\eta}=\vec{\alpha}/||\vec{\alpha}||$. Then, we can calculate

\begin{gather}
    \int^z_0 \hat{M}^{\mathrm{(I)}}_\mathrm{NL}(z') \mathrm{d}z'= \hbar g||\vec{\alpha}||\sum^N_{n,m,j=1} \eta_jS_{nj} S_{mj} \hat{b}^{\dagger}_n \hat{b}^{\dagger}_m 
    e^{i[\lambda_n+\lambda_m+
    \beta_j(\omega_\mathrm{p})]z/2}\mathrm{sinc}\{[\lambda_n+\lambda_m-
    \beta_j(\omega_\mathrm{p})]z/2\} 
    ze^{-i(\lambda_n+\lambda_m)z}
    + \mathrm{h.c.} \\
    = -i\hbar \sum^N_{n,m} \tilde Q_{nm} e^{-i(\lambda_n+\lambda_m)z}\hat{b}^{\dagger}_n \hat{b}^{\dagger}_m  + \mathrm{h.c.}\\
    =-i\hbar \sum^N_{n,m} \tilde Q_{nm} \hat{b}^{\mathrm{(I)}\dagger}_n \hat{b}^{\mathrm{(I)}\dagger}_m + \mathrm{h.c.},
\end{gather}
with

\begin{equation}
    \tilde Q_{nm}(z) = \delta(z) \sum_{j=1}^N\tilde P_{nmj}\tilde T_{nmj}(z), \label{Eq_SM_Qtilde}
\end{equation}

and
\begin{align}
    \delta(z) &= izg||\vec\alpha||, \\
    \tilde P_{nmj} &= \eta_jS_{nj} S_{mj},\\
    \tilde T_{nmj}(z) &= e^{i[\lambda_n+\lambda_m+
    \beta_j(\omega_\mathrm{p})]z/2},\nonumber\\
    &\times\mathrm{sinc}\{[\lambda_n+\lambda_m-
    \beta_j(\omega_\mathrm{p})]z/2\}. \label{Eq_SM_Tnmj}
\end{align}

Thus, the evolution in the interaction picture is given by

\begin{equation}
|\Psi(z)\rangle_\mathrm{I} = \mathrm{exp} \left[\frac{i}{\hbar} \int^z_0 \hat{M}^{\mathrm{(I)}}_\mathrm{NL}(z') \mathrm{d}z'\right]  |\Psi(0)\rangle_\mathrm{I},\label{Eq_SM_psi_ip}
\end{equation}

with the subscript $\mathrm{I}$ denoting that the states are in the interaction picture. Assuming a biphoton state, we can expand the exponential in Eq. (\ref{Eq_SM_psi_ip}) up to first order in $\eta_j$ and obtain

\begin{align}
    |\Psi(z)\rangle_\mathrm{I} 
    &\approx \left(\hat I + \frac{i}{\hbar} \int^z_0 \hat{M}^{\mathrm{(I)}}_\mathrm{NL}(z') \mathrm{d}z'\right)|\Psi(0)\rangle_\mathrm{I} \nonumber\\
    &=\left(\hat I + \sum^N_{n,m} \tilde Q_{nm} \hat{b}^{\mathrm{(I)}\dagger}_n \hat{b}^{\mathrm{(I)}\dagger}_m + \mathrm{h.c.}\right)|0\rangle_\mathrm{I} \nonumber\\
    &=|0\rangle_\mathrm{I} +\sum^N_{n,m} \tilde Q_{nm} \hat{b}^{\mathrm{(I)}\dagger}_n \hat{b}^{\mathrm{(I)}\dagger}_m|0\rangle_\mathrm{I}, \label{Eq_SM_Psi_IP_final}
\end{align}

with $\hat{I}$ the identity matrix. Switching back to the Schrödinger picture we arrive to

\begin{equation}
    |\Psi(z)\rangle=|0\rangle+\sum^N_{n,m} \tilde Q_{nm} \hat{b}^{\dagger}_n \hat{b}^{\dagger}_m|0\rangle,
\end{equation}

 which in the individual-mode basis reads 

\begin{equation}
    |\Psi(z)\rangle=|0\rangle+\sum^N_{j,k} Q_{jk} \hat{A}^{\dagger}_j \hat{A}^{\dagger}_k|0\rangle
    \label{Eq_solution_diff_betas}
\end{equation}
with

\begin{equation}
    Q = S^T\tilde Q S
\end{equation}

Let us consider the case with an injection only in waveguide $p$, such that Eq. (\ref{Eq_SM_Qtilde}) becomes $\tilde Q_{nm}(z) = \delta(z) \tilde P_{nmp}\tilde T_{nmp}(z)$. Hence, the effective propagation constant of the pump is only needed at waveguide $p$---that is, $\beta_p(\omega_\mathrm{p})$. For simplicity, we evoke a quasi-phase matching condition---for example, due to periodic poling. Thus, we can replace $\beta_p(\omega_\mathrm{p})$ by $\tilde{\beta}_p\equiv\beta_p(\omega_\mathrm{p})+\beta_\mathrm{QPM}$ in the expression for $\tilde T_{nmp}$, with $\beta_\mathrm{QPM}$ the required quasi-phase matching contribution to achieve $\tilde{\beta}_p=2 \beta_p(\omega_s)$.

\newpage

\subsection{Border effects}

\begin{figure}[H]
    \centering
    \includegraphics[width=0.75\linewidth]{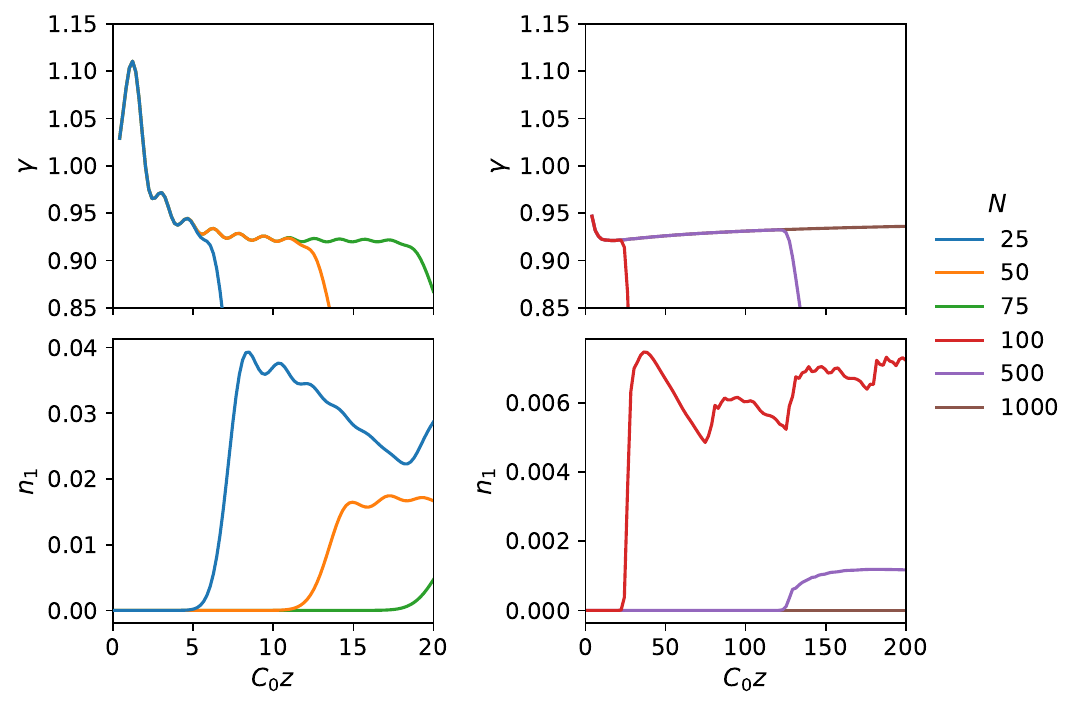}
    \caption{Border effects for an injection in the center waveguide of ordered ANWs with $N$ identical waveguides ($\beta_j=\beta_0$) and homogeneous coupling profile ($C_j=C_0$). The upper, and lower rows depict the exponent $\gamma$, and the  normalized photon number of the corner waveguide with index 1 ($n_1$), respectively. The cases corresponding to arrays of 25, 50 and 75 waveguides are portrayed in the first column, whereas those belonging to arrays of 100, 500 and 1000 waveguides are shown in the second column. In the upper-right panel, the initial maximum of $\gamma$ seems to be absent due to the rendering.}
    \label{fig:size_effect_center}
\end{figure}

\begin{figure}[H]
    \centering
    \includegraphics[width=0.75\linewidth]{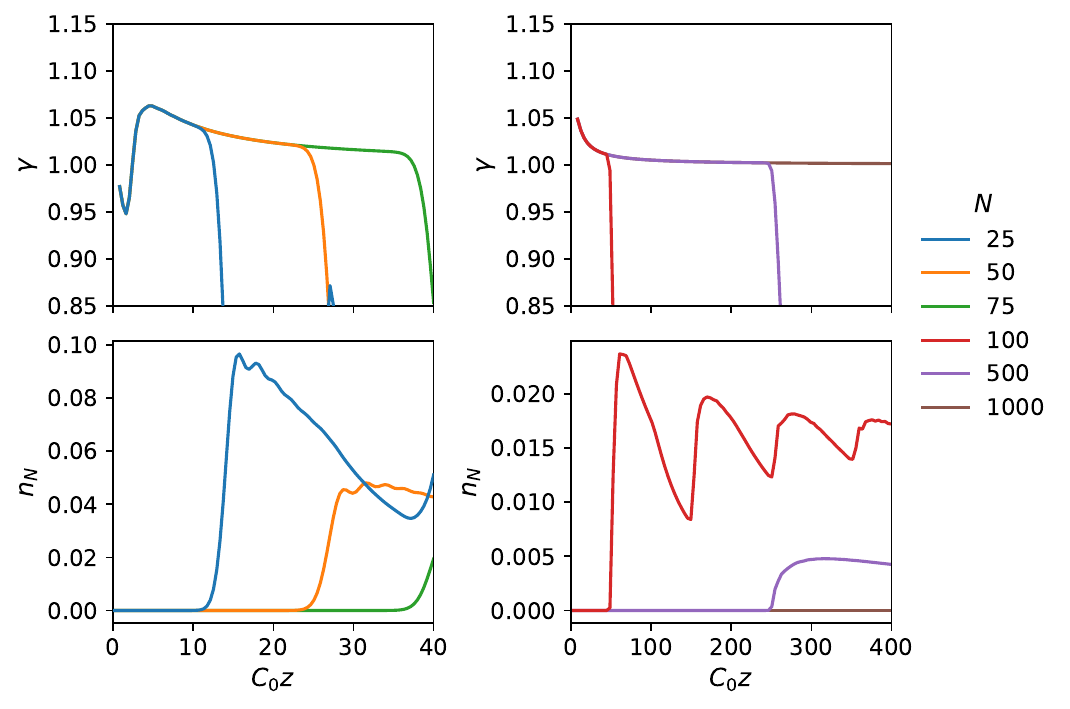}
    \caption{Border effects for an injection in the corner waveguide of ordered ANWs with $N$ identical waveguides ($\beta_j=\beta_0$) and homogeneous coupling profile ($C_j=C_0$). The upper, and lower rows depict the exponent $\gamma$, and the  normalized photon number of the corner waveguide with index $N$ ($n_N$), respectively. The cases corresponding to arrays of 25, 50 and 75 waveguides are portrayed in the first column, whereas those belonging to arrays of 100, 500 and 1000 waveguides are shown in the second column. In the upper-right panel, the initial minimum of $\gamma$ seems to be absent due to the rendering.}
    \label{fig:size_effect_corner}
\end{figure}

Figures \ref{fig:size_effect_center}, and \ref{fig:size_effect_corner} depict the border effects observed in ordered and homogeneous ANWs (with $C_j=C_0$ and $\beta_j=\beta_0$) of different size $N$ for an injection in the center, and corner waveguide, respectively. These can be identified by a steep increase in $n_1$ and a concomitant sharp decrease of $\gamma$. As expected, $n_1$ and $\gamma$ are indistinguishable within regions without border effects for any size $N$. It is of interest to obtain the value of $\gamma$ corresponding to a large propagation distance without border effects. Particularly, for an injection in the center waveguide with $N=1000$, $\gamma$ reaches a value of 0.936 at $C_0z=200$. Also, for an injection in the corner waveguide, a value of $\gamma=1.001$ is observed at $C_0z=400$ in an array with $N=1000$.

\begin{figure}[H]
    \centering
    \includegraphics[width=0.70\linewidth]{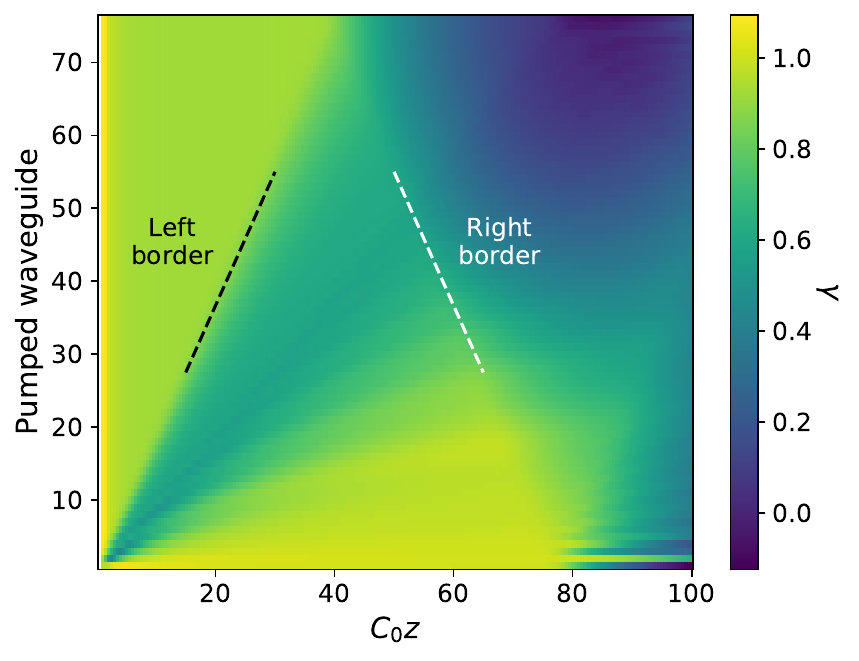}
    \caption{Dependence of $\gamma$ with respect to the propagation distance as the pumped waveguide is varied in an ordered and homogeneous ANW of size $N=151$. Let the indexes $1$, and $N$ correspond to the left, and right border, respectively. The straight lines depicted in the figure indicate the propagation distances upon which the border effects from the left and right borders become relevant.}
    \label{fig:border_effect}
\end{figure}

Figure \ref{fig:border_effect} depicts $\gamma$ as a function of the propagation distance and injected waveguide in an array of $N=150$. It provides an example of the behavior of $\gamma$ in cases in which the injected waveguide is neither the center nor the corner one. Interestingly, it is possible to identify the propagation distances corresponding to the decrease in $\gamma$ due to the beginning of the effects assigned to both the left (waveguide $1$) and right (waveguide $150$) borders, as depicted in Figure \ref{fig:border_effect} by straight lines.

